%% file: general_overview_of_formalisms_for_individual_based_modeling.tex
\documentclass[11pt,a4paper]{article}
\usepackage{amsmath}
\usepackage{amsfonts}
\usepackage{amssymb}
\usepackage{mathptmx}
\usepackage{float}
\usepackage{graphicx}
\usepackage{mdwlist}
\usepackage{graphics}
\usepackage[usenames]{color}
\usepackage{hyperref}
\usepackage{listings} 
\usepackage{amssymb} 
\usepackage{amsthm}

\usepackage{times}
\usepackage{amssymb}
\usepackage{amsmath}
\usepackage{latexsym}
\usepackage{graphicx}

\usepackage[strings]{underscore} 
\usepackage{mdwlist}
\usepackage{subfigure}
\usepackage{wasysym}
\usepackage{amsthm } 
\usepackage{algorithm}
\usepackage{algpseudocode}

\input{macros}

\begin{document}
\title{A General Overview of Formal Languages for Individual-Based Modelling of Ecosystems}
 \author{Mauricio Toro\\Universidad Eafit\\mtorobe@eafit.edu.co}
\date{ }

\maketitle

\section*{Abstract}
Various formal languages have been proposed in
the literature for the individual-based modelling of ecological systems.
These languages differ in their
treatment of time and space. Each modelling language offers a distinct view and
techniques for analyzing systems. Most of the languages are based on process calculi or
P systems. In this article, we present a general overview of the existing modelling languages based on process calculi. We also discuss, briefly, other approaches such as P systems, cellular automata and Petri nets. Finally, 
we show advantages and disadvantages of these modelling languages and we propose some future research directions.

\section{Introduction}

The collective evolution of a group of individuals is of importance
in many fields; for instance, in system biology, ecology and epidemiology.
When modelling such systems, we want to know the emergent behavior of the
whole population given a description of the low-level interactions of the individuals
in the system. As an example, in \emph{eco-epidemiology} the focus is on the evolution in time of the number of individuals
infected in a certain population  and how a small number of individuals infected may lead to an epidemic.

Eco-epidemiology can be seen as a particular case of \emph{population ecology}.
The main aim  of population ecology is to gain a better
understanding of population dynamics and make predictions about how
populations will evolve and how they will respond to
specific management schemes. In epidemiology, such management schemes can be
 a cure to a disease, mechanisms to prevent a disease such as vaccines, or
mechanisms to prevent the \emph{vector} (species infected with a disease) to spread a disease.
To achieve these goals, scientists construct models of ecosystems and management schemes (e.g., \cite{Yang2008}).

Luisa Vissat et al. argue that we can use these models to
predict the system behaviour and use this predictive power to support decision-making  \cite{mela}. As an example, in
the case of the spread of diseases, models are used to find the optimal control strategies for containing
the spread or to predict the results of a population reduction, as a key way
to control disease in livestock and wildlife \cite{wildlife}.

Various formalisms have been proposed in
the literature for the individual-based modelling of ecological systems.
These formalisms differ in their
treatment of time and space and each modelling language offers a distinct view and
techniques for analyzing systems. Surprisingly, many formalisms developed for population modelling have been also used to model music interaction in real life performance; for instance process calculi has been used to model real-time performance in music \cite{toro2015ntccrt}, computer music improvisation \cite{toro2016} and interactive music scores \cite{tdcr14}. In fact, process calculi have been extensively applied to the modeling of interactive music systems
 \cite{toro2019b, toro2019, toro2018h, toro2018g, toro2018f, toro2018e, tdcjo2018, toro2018d, toro2018c, toro2018b, toro2018, toro2016, toro2016faust, toro2016gelisp, 2016arXiv160202169T, toro2015ntccrt, is-chapter,tdcr14,ntccrt,cc-chapter,torophd,torobsc,Toro-Bermudez10,Toro15,ArandaAOPRTV09,tdcc12,toro-report09,tdc10,tdcb10,tororeport} 
 and ecological systems \cite{EPTCS2047, PT13,TPSK14,PTA13}. In addition, research on algorithms \cite{PAT2016, MorenoPT17, RestrepoPT17, qsrt2018, rhrat2018}
 and software engineering \cite{tsmvvcrwc2018,toro2018a} also contributes to this field.

In what follows, we explain, in Section 2, classic process calculi and different extensions of classic process calculi used to model population systems and some applications in ecology; in Section 3, P systems and some extensions used to model population systems; and, in Section 4, other approaches used for the formal modelling of population systems. Finally, in Section 5, we discuss the advantages and disadvantages of these formalisms and we recommend some future research  directions in this field. Note that the formalisms explained in this article are not only strictly related to applications in ecology. Futhermore, this paper focuses on process calculi and P systems, but we also present other 
approaches for formal modelling of population systems such as cellular automata and Petri nets.

\section{Process calculi}
 \emph{Process calculi (or process algebras)} are a diverse family of related approaches for formally modelling concurrent systems. Process calculi provide a tool for the high-level description of interactions, communications, and synchronizations between a collection of independent processes. Process calculi also provide algebraic laws that allow process descriptions to be manipulated and analyzed, and allow formal reasoning about equivalences between processes.

In what follows, first we explain classic process calculi. After that, we explain, in more detail, different extensions of classic process calculi used to model population systems and some applications in ecology. Note that process calculi explained in this section are not only strictly related to applications in ecology.

\subsection{Classic Process Calculi}

\paragraph{CCS. }
The \emph{calculus of communicating systems} (\ccs) \cite{Milner-CCS} is a
process calculus 
introduced by Robin Milner around 1980. The calculus includes
primitives for describing parallel composition, choice between actions and scope
restriction. \ccs{} is useful for evaluating  qualitative correctness 
properties of a system such as deadlock or livelock. 
The expressions of the language are interpreted as a \emph{labeled transition
system}.
Between two labeled transition systems, \emph{bisimilarity} is used as a
semantic equivalence.
 In fact, there is a tool to check bisimulation, as well as for the simulation and
model checking
of \ccs{} and \sccs{}
\footnote{\url{http://homepages.inf.ed.ac.uk/perdita/cwb/doc/}}.



\paragraph{SCCS. }
Soon after the creation of \ccs{}, Milner developed an extension of \ccs{}. In
\emph{synchronous \ccs{}} (\sccs{}) \cite{Milner-SCCS}, concurrent processes
proceed, in
contrast to \ccs{}, simultaneously in
lockstep. This means that at every step the concurrent processes in a composite
system
perform a
single action. 

\paragraph{CSP. } 
\emph{Communicating sequential processes} (\csp{}) \cite{Hoare-CSP} is a process
calculus first described in 1978 by  C. A. R. Hoare. 
\csp{} is different from \ccs{}: In \csp{}, processes
composed in parallel cooperate in a multi-party synchronization. In contrast,
in \ccs{}, 
parallel processes interact with each other by means of a two-party
synchronization. 
Another difference between \csp{} and \ccs{} is that
synchronization in \csp{} is made by symmetrical actions, whereas in \ccs{} there
are actions and co-actions. 
Soon after the invention of \ccs{} and \csp{}, the need of sending channel
names to channels themselves motivated the invention of the pi-calculus.

\paragraph{Pi-calculus. }
The \emph{pi-calculus} (or $\pi$-calculus) \cite{Milner-PI} is a process
calculus developed by Robin Milner, Joachim Parrow and David
Walker in 1992, as a continuation of Milner's work on \ccs{}. The pi-calculus
allows
channel
names to be communicated along the channels themselves. Using this feature, one
is 
able to describe concurrent computations whose network configuration may change
during the computation.

\paragraph{CCP.} \emph{Concurrent constraint programming} (\textsc{ccp}) was
developed by Vijay Saraswat in 1991. In \ccp{}, a system is modeled 
in terms of processes adding to a common \textit{constraint store} partial
information on
the value of variables. Concurrent processes synchronize by blocking until a
piece of information can be deduced from the store. \textsc{ccp} is
based on the idea of a \textit{constraint system (\textsc{cs})}. A constraint
system is composed of a set of (basic) constraints and an entailment relation
specifying  constraints that can be deduced from others.

\subsection{Stochastic process calculi}

Stochastic process algebras aim to
combine  two successful approaches of modelling: 
\emph{labeled transition systems} and \emph{continuous-time Markov chains} \cite{NLLM09}.
Indeed, they argued that labeled transition systems are a very convenient 
framework to provide compositional semantics
of concurrent languages and to prove qualitative properties. Markov chains,
instead, have been used in
performance evaluation and quantitative properties. 
The common feature of most stochastic process calculi is that their actions are
enriched with rates of exponentially distributed random variables
characterizing their mean duration.

\paragraph{Continuous Pi-Calculus. } Ever since the invention of the pi-calculus,
there has been an interest to use it to model  ecosystems. A successful
development on this direction is an extension of the pi-calculus with continuous
time. 
The \emph{continuous pi-calculus}  \cite{KS08} is an extension of the
pi-calculus in which
the processes evolve in continuous time. It can be seen as modular way to
generate ordinary differential equations for a
system behavior over time. The continuous pi-calculus  has been used to model biochemical 
systems  \cite{KS08}.

\paragraph{Stochastic \ccs. }
In \emph{stochastic \ccs} (sto\ccs), output actions are equipped
with a parameter
characterizing a random variable with a negative exponential distribution,
modelling the duration of the action \cite{NLLM09}. Input actions are denoted
with a weight,
a positive integer used to determine the probability that the
specific input is selected when a complementary action is executed \cite{NLLM09}. There are no tools available for
their simulation nor verification up to our knowledge. This calculus has
not been used to model ecosystems, but the definition of \emph{rate-based transition
systems} has inspired many other modelling languages \cite{NLLM09}.

As argued by Cardelli et al. in \cite{CM12}, one cannot guarantee associativity
of parallel composition
operator up to stochastic bisimilarity when the synchronization paradigm of \ccs
is used in combination with the synchronization rate computation based on
apparent rates. This is a
problem, specially, in presence of dynamic process creation. 

\paragraph{Stochastic pi-calculus. }
The \emph{stochastic pi-calculus} is an extension of the pi-calculus
with rates associated to the actions, developed by Corrado Priami in 1995.
Recently, Cardelli et al. developed new
semantics with replication and fresh name
quantification \cite{CM12}. Cardelli et al. argued that 
parallel composition failed to be associative up to stochastic 
bisimilarity with previous semantics. With the new semantics  it is now
possible to capture associativity, which will lead to new applications and
simulation tools. In fact, there is already a simulation tool for the stochastic
pi-calculus named the \emph{stochastic pi
machine (Spim)} 
\footnote{\url{http://research.microsoft.com/en-us/projects/spim/}}; Spim 
uses the probabilistic model checker \prism{}. Another
approach for model checking has been
developed by Norman et al. \cite{NPPW09}. 

Stochastic pi-calculus has been used in system biology to model regulating gene expression by positive feedback \cite{Phillips04acorrect}.

\paragraph{PEPA. }
The
properties that may be checked for concurrent systems modeled by an algebraic
description 
include freedom from deadlock and
algebraic equivalence under observation \cite{GH94}. Hillston et al. argue that
there are further properties of interest for process calculi models 
such as steady-state probabilities and rewards for performance
measures. \emph{Performance evaluation process algebra} (\pepa{}) \cite{GH94}
was defined by Hillston et al., in 1994, to allow the specification and
verification of such 
properties. Every activity in \pepa{} has an associated duration which is a
random
variable with an exponential distribution, determined by a single real-number
parameter. 
A \pepa{} process can be translated into a continuous-time Markov process
\cite{GH94}.

\pepa{} has been used in system biology to model the \emph{raf kinase inhibitor protein} on the extracellular signal regulated kinase signalling pathway \cite{Hillston2006} and to model measles epidemics in the UK from 1944--1964 \cite{measles2012}.

\paragraph{Bio-\pepa{}. }
\emph{Bio-\pepa{}} \cite{CH09} is a process calculus for modelling and analyzing
biochemical
networks. Bio-\pepa{} is a modification of \pepa{} adding some features such as support
for
\emph{stoichiometry} and general \emph{kinetic laws}.
 Kinetic laws are functions which allow us to derive the rate of reactions from
varying parameters such as the rate coefficients and concentration of the
reactants 
\cite{Pardini-thesis}. 
According to Hillston et al., the main difficulty with \pepa{} is the
definition of stoichiometric
coefficients \cite{CH09}: These coefficients are used to show the quantitative
relationships of the reactants and products in a biochemical reactions. 
A major
feature of Bio-\pepa{} is the possibility to represent explicitly some features
of biochemical models such as stoichiometry and the role of species in a given
reaction. 
 Bio-\pepa{} is also enriched with some notions
of equivalence such as isomorphism and bisimulation, extended from
\pepa{}. 
Bio-\pepa{} has been used in epidemiology to model the \emph{H5N1 avian influenza} \cite{CiocchettaH10}.

\paragraph{Stochastic CCP. } Stochastic \ccp{} (s\ccp{}) was developed by
 Bortolussi \cite{Bortolussi06}, in 2006. s\ccp{} is an extension of \ccp{} by adding a
stochastic duration to all instructions interacting with the constraint store.
In s\ccp{}, each instruction has an associated random variable, exponentially distributed \cite{Bortolussi06}.
Bortolussi et al. also proposed an approximation of s\ccp{} semantics into  ordinary differential equations \cite{BortolussiP07}.

Bortolussi et al. have used s\ccp{} to model
bio-mechanical reactions such as an \emph{enzymatic reaction}  \cite{Bortolussi06}. 
s\ccp{} has also been in system biology \cite{Bortolussi2008} and to model prey-predator dynamics \cite{BortolussiP07}.

\subsection{Probabilistic Process Calculi}

Process calculi depend on concepts of non-deterministic choice. Tofts argues
that it is natural to
consider extending such systems by adding
a probabilistic quantification for non-determinism \cite{Tofts94}.
In terms of expressiveness
of modelling, McCaig et al. argue that probabilistic choice is a more natural
way to express
individual behavior than stochastic rates of activities for epidemiological
models \cite{MNS10}. Another alternative for semantics of probabilistic calculi is having both non-deterministic and probabilistic choice, 
as explained by Norman el at. \cite{NPPW09}.

\paragraph{Weighted SCCS. }

A probabilistic calculus, derived from Milner's \sccs{} \cite{Milner-SCCS} , is \emph{weighted 
\sccs{}} (\wsccs{}) \cite{Tofts94}. This calculus was developed by Chris Tofts, 
in 1994. \wsccs{}
is
synchronous because the purpose is to quantify the
relative
frequency of free simultaneous choice. In an asynchronous system, 
choices may be resolved at arbitrary times, thus giving a choice which may not
be
between equally free objects. \wsccs{} does not  quantify choices directly
with probabilities, but it uses
weights: Weights are interpreted as probabilities via the concept
of relative frequency.

McCaig et al. presented a new semantics in terms of 
\emph{mean field equations} 
in \cite{MNS10}. 
This semantics captures the average behavior of the system over time, without
computing the entire state space; therefore, avoiding the state-space explosion
problem. 
The new semantics is shown to be equivalent to the standard discrete-time Markov
chain semantics of \wsccs{}, as the number of processes tends to infinity.
Using \emph{\wsccs{}'s probabilistic
workbench}, one 
can analyze systems in \wsccs{} \cite{MNS10}.  There is also some work on population models for stochastic process algebras done by Hillston et al.
    \cite{HillstonTG12,Tribastone12}. A related interesting work is the model checking of mean-field approximations of
continuous-time Markov chains
\cite{BortolussiHLM13}.

Note that this calculus has been used to
model human-population growth \cite{MNS10} and \wsccs{} has been employed in various ecological studies
by its author and others~\cite{SBB01,MNS08}.

\paragraph{Probabilistic \csp\#. }

\emph{Hierarchical probabilistic \csp{}} (\textsc{pcsp}\#) \cite{SSL10} extends
\csp{}
with
probabilistic choice and data structures. \textsc{pcsp}\# was developed by Jun
Sun et al. in 2010. 
\textsc{pcsp}\# combines sequential programs defined in a simple imperative
language such
as
C\# with high-level specifications such as parallel composition, choice and
hiding, as well
as probabilistic choice. Its underlying semantics is based on \emph{Markov
decision
processes}.

Although \textsc{pcsp}\#'s semantics are given on Markov decision processes, Sun
et al. argue that existing probabilistic model checkers have been designed
for simple systems without hierarchy \cite{SSL10}.
For that reason, they extended an existing model checker to support
probabilistic model checking
of hierarchical complex systems.  
Probabilistic \csp{} has been used to model sensor networks, security protocols
and probabilistic algorithms. \textsc{pcsp}\# has not been used to model ecosystems.

\paragraph{Probabilistic pi-calculus. }
The \emph{probabilistic pi-calculus} was developed by Herescu et al. in
2000 \cite{Herescu2000}. 
In the probabilistic pi-calculus, there are two types of
choice operators: non-deterministic and probabilistic. 
The operational semantics for probabilistic extensions of
the pi-calculus are typically expressed in terms of Markov decision
processes. Existing
semantics are concrete, which means processes are encoded directly into Markov decision
process. Norman et al., defined a symbolic representation of the calculus
semantics \cite{NPPW09}. 
The main feature of Norman et al.'s symbolic semantics is that the input
variable of input
transitions is kept as a name variable and the communication rule that matches
between the input and the output channel is represented by a condition
\cite{NPPW09}. Norman et al. used 
those semantics for model checking of the calculus using probabilistic model checker \prism{}. 
Norman et al. also used this calculus to model and verify randomized security protocols \cite{NPPW09}.
Probabilistic pi-calculus has also been used for model checking models of system biology \cite{Norman2008}.

\paragraph{Probabilistic CCP.} There is a timed, probabilistic,
non-deterministic extension of \ccp{} (\textsc{pntcc}) \cite{pntcc}, developed
by P\'erez et al. in 2009. They defined the semantics as a discrete-time Markov
chain in which non-determinism is unobservable because it is removed by a
scheduler that must be defined along with the model. \textsc{pntcc} has not been used for models
in system biology nor ecology.

\subsection{Spatial Process Calculi}
Spatial aspects and locations are required in ecosystems because such 
systems can be physically distributed in space and this distribution can affect
the length of time required for an activity to occur \cite{Galpin09}. Spatial
process calculi are extensions of process calculi with different notions of
space. Spatial features of biological systems can be
studied in different ways \cite{Pardini-thesis}. Pardini argues that most
computer-science formalisms focus on the  abstract spatial feature
which concerns the \emph{modelling of compartments}; nonetheless, he also argues
that
there are other formalisms that focus on a more \emph{concrete notion of space}
(e.g., continuous 2D space) \cite{Pardini-thesis}.

\paragraph{\palps{}. }

\emph{Process algebra with locations for population systems (\palps{})} can be considered as an extension of \ccs{} with probabilistic choice,
and
with 
locations
and location attributes \cite{PTA13}.  It shares a similar treatment of locations with
process
algebras developed for reasoning about mobile ad hoc networks; for
instance, Galpin's spatial extension to \pepa{} and
Kouzapas et al.'s calculus for dynamic networks \cite{KP11}. Nonetheless, such
approaches are stochastic. \palps{} shares some similarities with \sccs{} 
 although it does not provide synchronous parallel composition nor
probabilistic choice based on weights. 
\palps{} considers a two-dimensional space where locations and their
interconnections are modeled
as a graph upon which individuals may move as computation proceeds. 
 Compartments are static in \palps{}. In this matter, \palps{} is related to the
spatial calculi previously described
such as the calculus of wrapped compartments or to cellular automata.
\palps{} was used to model prey-predator dynamics \cite{PTA13}.

\palps{} was also extended with \emph{process ordering}  because simulations carried out by ecologists often impose an order on
the events that may take place within a model \cite{PT13}. As an example, if we
consider mortality and reproduction within a single-species model,
three cases exist: concurrent ordering, reproduction preceding
mortality and reproduction following mortality. 
Process ordering in \palps{} was used to model the reproduction of the \emph{parasitic
varroa mite} that attacks honey bees \cite{PT13}.

\paragraph{Synchronous \palps{} (\spalps{}). }

To alleviate
 the problem of the interleaving nature of parallel composition and high level of non-determinism in \palps{}, Toro el al.  proposed  a new semantics of \palps{}, and an associated \prism{} translation, 
that disassociates the number of modules from the maximum number of individuals, 
while capturing more faithfully the synchronous evolution of populations and
removing as much unnecessary non-determinism as possible. The synchronous extension of \palps{} is named \emph{synchronous \palps{} (\spalps{})} \cite{TPSK14}. This synchronous semantics of \palps{} implements
the concept of \emph{maximum parallelism}: at any given time all individuals
that may execute an action will do so simultaneously.

Toro et al. developed a mean-field semantics for \spalps{} \cite{EPTCS2047}.
This semantics allows an interpretation
of the average behavior of a systems as a set of recurrence equations.
 Recurrence equations are an approximation  useful when dealing with a large
 number of individuals, as it is the case of epidemiological studies.  
\spalps{} was used to model dengue transmission in Bello (Antioquia), Colombia, in particular, to analyze the factors involved in the epidemiology of dengue disease \cite{EPTCS2047}. The model presented by Toro et al. in \cite{EPTCS2047} is an individual-based version of the model presented in \cite{Yang2008}.

\paragraph{Spatial  \pepa{}. }
Galpin introduced spatial notions to the stochastic process algebra \pepa{},
creating \emph{spatial \pepa{}} \cite{Galpin09}, in 2009.
Galpin was motivated by both computer networks and epidemiology where locations
of
actions or processes may affect the time taken by an event. She defined a very
general
stochastic process algebra with locations, where locations are introduced
to both actions and processes. 

Galpin defined a weighted hypergraph structure to represent locations. The 
locations may or
may not have structure. Nodes in a directed graph may 
represent computers in a
network or locations in n-dimensional space. The graph edges could represent
simple network connectivity, or a tree structure representing nesting of
locations. As another example of its
generality, the weight in the hypergraph structure could be obtained from
distance function or from something
else.

Galpin used Spatial  \pepa{} to model single packet traversal through network \cite{Galpin09}. 
Spatial  \pepa{} has not been used to model ecosystems. 

\paragraph{BioAmbients. }

\emph{BioAmbients} \cite{RPSCS04} calculus was developed by Cardelli et
al. in 2004. It was motivated by the Ambient calculus \cite{ambient}, originally developed by
Cardelli 
for the specification of process location and movement through computational
domains \cite{ambientCardelli}. 
Bio-molecular systems, composed of networks of proteins, underlie the major
functions of living cells. Compartments are the key to organization of such
systems, according to Cardelli et al., BioAmbients is inspired in such biological
compartments. The novelties with respect to the Ambient calculus are the
movement of molecules between compartments, the
dynamic rearrangement of cellular compartments and the interaction between
molecules in a compartmentalized setting. Cardelli et al. adapted the
\emph{BioSpi} simulation tool 
for their new calculus and used it to study a complex multi-cellular system \cite{RPSCS04}.

\paragraph{Brane calculus. }
A biological cellular membrane is an oriented closed surface that can perform
various molecular functions. One constraint of such membranes is bi-tonality.
 This constraint 
requires nested membranes to have opposite orientations, so they can be coded
with a coloring system of two tones. The organization of these membranes
inspired the creation of the \emph{Brane calculus}, in 2004, by Cardelli et al.  
\cite{Cardelli-BC}. In
Brane, there is composition of systems,
composition of membranes and replication. 
There is also a choice operation that can be added to membranes and
communication on 
\ccs{} style.
 Cardelli argues that it is difficult to encode this calculus on BioAmbients
although they are 
 closely related. The Brane calculus has been used to model the behavior of
 the \emph{semliki forest virus} \cite{Priami2006}.

\paragraph{Calculus of wrapped compartments. }

The \emph{calculus of wrapped compartments} (\textsc{cwc})  \cite{BCCDSPT11} is
 also based on the notion of a
compartment. 
\textsc{cwc} 
has no explicit structure modelling a
spatial geometry; nonetheless, Bioglio et al. argue that the compartment
labeling feature can be exploited to model various examples of spatial
interactions in a natural way  \cite{BCCDSPT11}. 

To model 2D-space, \textsc{cwc} allows to label a compartment
as a 2D-coordinate,
denoted by a tuple of integers (row,column), thus they can define a
two-dimensional grid
\cite{BCCDSPT11}. Modelling the space as a two-dimensional grid, 
this calculus was used to analyze the growth model for
\emph{arbuscular mychorrizal fungi} \cite{BCCDSPT11}. 
Biochemical transformations are
described via a set of stochastic reduction rules which characterize the
behavior of the represented system. 
Stochastic simulation can be defined along the lines of Gillespie's simulation. Bioglio et al. performed simulations of the growth model using the
\emph{surface
language} (a language they developed) \cite{BCCDSPT11}.

\paragraph{Spatial pi-calculus. }
Most calculi have focused on population-based approaches or on how to
place and
move individuals relative to each others position in space. \emph{SpacePi}
\cite{JEU07}
extends the pi-calculus 
with continuous time and space. Processes are embedded into a
vector
space and move
individually. Only processes that are sufficiently close can communicate. 

According
to John et al., existing
spatial individual-based approaches have so far focused on indirect,
relative space \cite{JEU07}. For that reason, John et al. 
developed SpacePi to explicitly model location and movement of individuals in 
absolute space. 
By combining an individual-based perception and absolute space, not only
diffusion processes, but also active transportation inside membranes can
be modeled realistically at different resolutions \cite{JEU07}. 
This makes their approach
different from other pi-calculus extensions, like the stochastic pi-calculus. This implies
that existing simulation engines based on Gillespie's algorithm cannot be used; 
this also implies, that simulation will be less
efficient than the one obtained with engines based on Gillespie's algorithm. 
This calculus has been used by John et al. to model the behavior of a euglena, a
micro-organism living in inland water which motion is influenced by sun
irradiation. John et al. ran simulations on a tool of their own \cite{JEU07}.

\paragraph{Spatial \ccp{} . } Barco et al. recently proposed a spatial extension of
\ccp{} \cite{KPPV12}. In spatial \ccp{}, each process has a computation
space that can be shared with other processes. In
addition, it is possible to have nested computation spaces. Spatial \ccp{} has not
been used to model ecosystems.

\paragraph{Modelling in Ecology with
Location Attributes. }
 \emph{Modelling in Ecology with
Location Attributes (\textsc{mela})} is a process calculus to formally describe ecological systems, with focus on space abstraction and environmental description \cite{mela}. Like other stochastic process algebras, \textsc{mela} allows complementary analysis
techniques, such as stochastic simulations, numerical solution of ordinary differential equations. In \textsc{mela}, the modeller can choose between different spatial structures; for instance, a graph, a
discrete line segment, a 2D grid, a 3D grid, or a nested spatial structure. \textsc{mela} has been used to model 
 prey-predator dynamics (the Lotka-Volterra model) and the spread of a Cholera epidemic \cite{mela}. A disadvantage of \textsc{mela}
 is the lack of tools to support the spatial and temporal analysis that could be possible using its semantics.

\paragraph{Process Algebra for Located
Markovian Agents.} \emph{Process Algebra for Located
Markovian Agents (\textsc{paloma})} \cite{paloma} uses multi-class  multi-message  Markovian  agents  to model collective systems comprised of populations of agents which are spatially distributed. \textsc{paloma} is equipped with both discrete event and differential semantics.
The discrete semantics provides
the theoretical foundation for discrete event simulation and the differential semantics allows us to automatically derive the underlying mean-field model. \textsc{paloma} was used to  model  a simplified scenario of the 1918-1919 flu epidemic in central Canada and
to investigate the effect of quarantine on the spread of the flu.

\paragraph{Other extensions of CCS. } Bartocci et al. proposed an extension of CCS to model 3D space.
This extension is called the \emph{shape calculus}   \cite{BartocciCBMT10, BartocciCBMT10a}.

\paragraph{Other extensions of the Pi-calculus. }  Pardini described several spatial extensions of the Pi-calculus
\cite{Pardini-thesis}: He mentioned \emph{Beta binders}, $\pi@$ calculus,
\emph{attributed $\pi$-calculus}, the \emph{imperative pi-calculus} and the
\emph{$3\pi$ calculus}. In Beta binders, the notion of compartments is represent
by the
means of \emph{boxes} which contain pi-calculus processes. In $\pi@$ calculus,
channels are associated with compartments. In attributed pi-calculus, processes
can have attributes, for instance, information of their location. In
imperative pi-calculus, there is a global store with information on the volume
of each compartment. Finally, in the $3\pi$ calculus, processes are embedded in
3D space. 
 It is worth to note that some of these extensions have been also extended with
stochasticity, as it is the case of the $\pi@$ calculus.

Beta binders has been used to model biological interactions \cite{Priami2005} and
the $\pi@$ calculus has been used to model the \emph{euglena’s phototaxis}, 
gene regulation at \emph{lambda switch} and population models \cite{John2010}.

\section{P systems}

A different approach towards modelling of ecological
systems is that of  \emph{P systems} \cite{Paun01, MC}. 
\emph{P system} \cite{Paun01} is a class of distributed and parallel computing
described
by Gheorghe P\u{a}un  in 2001. Each \emph{membrane} represents a region
and contains a
multiset of objects.
A \emph{cell} is considered, in an abstract way,  a hierarchy of
compartments enclosed
by membranes. Each compartment may include elementary objects as well as other
compartments. Processes in a cell are viewed as sequences of discrete events.

 Romero et al. argue that P systems started from the
 assumption that the processes
 taking place in the compartmental structure of a living cell can be
interpreted
 as
 computations \cite{RGBPPC06}.  Romero et al. argue that a P system consist of cell-like
 membrane structures (known as 
 compartments) in which one places multisets of objects that evolve according
to
 given rewriting rules (i.e., evolution rules).

\subsection{Stochastic P systems}


Spicher et al. described a sequential strategy for executing P systems based on
Gillespie's
stochastic simulation
algorithm \cite{SMCGP06}. They defined a stochastic simulation of
 a new class of P systems. This new kind
of P systems are called \emph{stochastic P systems}. 
In rewriting systems, the policy for deciding which
rule(s) will be applied is
called 
the application strategy \cite{SMCGP06}. They argued that in case of conflicts,
rules are selected
non-deterministically. Maximal-parallel
strategy
is well suited for the
modelling of discrete dynamic systems operating synchronously. This strategy 
means that the rules are applied
simultaneously in a maximal way during each rewriting step. This strategy is
less 
suited to capture events that occur asynchronously in continuous time.

Gillespie's algorithm is an alternative rule application strategy
for P systems.
Gillespie's algorithm leads to a sequential application strategy for these
rules: only one rule is applied in each derivation (simulation) step \cite{SMCGP06}.

Spicher et al. implemented stochastic P systems using the
spatially-explicit
programming language MGS \cite{SMCGP06}. Spicher et al.  have used stochastic P
systems to model the \emph{Lotka-Volterra
auto-catalytic system} and the life cycle of the \emph{similiki forest virus} \cite{SMCGP06}.
Romero et al., provide a translation of stochastic P systems to \prism{}, 
transforming a P system specification into a \prism{} input \cite{RGBPPC06}.
Romero et al.'s translation allows model checking of stochastic P
systems.



\subsection{Probabilistic P systems}

Probabilistic P systems
have been applied to a wide range of situations  in the field of
ecology (e.g., ~\cite{three}). 
 The semantics of probabilistic P systems is closely related to
 \spalps{}: rules are usually applied with maximal parallelism while
 several proposals have been considered on resolving the non-determinism that
 may arise when more than one combination of rules is possible, for instance,~\cite{PBMZ05,CCMPPPS11}.
 Probabilistic P systems have been applied to model the population
 dynamics of various ecosystems~\cite{BCPM06,abs-1008-3301}
 as well as to study the invasiveness of eastern species in European water frog populations~\cite{Pasquale}.
 As an
example, there is a model of mosquitoes in Italy that uses attributes such as temperature and humidity 
 \cite{abs-1008-3301}.

In \emph{probabilistic P systems}, the rules
include a
real number between zero and one: a probability of
the rule for its application \cite{CCMPJD09}. As expected, the sum of all
 rules  with the same left-hand
side of a rule must be one. The rules are applied in a maximal-consistent
parallel way, as usual. 
The inherent randomness
and uncertainty in ecosystems is captured by using probabilistic strategies.

Another application of probabilistic P systems in ecology is to model
 an ecosystem of some scavenger birds \cite{CCMPJD09}.
 Cardona et al. described a P system for modelling an ecosystem based on three
vulture
species and the prey species  from which vulture species obtain most of
their
energy requirements \cite{CCMPJD09}.
They include rules to limit the maximum amount of animals that can be
supported by the ecosystem as well as the amount of grass available for the
herbivorous species. They consider two regions: the skin and the
inner membrane. The first region controls the densities of every species
do not overcome the threshold of the ecosystem. 
The skin is where the animals reproduce and the inner membrane where they feed
or die. Note that this model is not spatially explicit. 
 Cardona et al.  simulated their ecosystem model using the  specification language
\emph{P-Lingua} for P
systems \cite{CCMPJD09}.

\subsection{Other extensions of P systems. } There are at least two extensions to
probabilistic P systems: \emph{Dynamical probabilistic P systems} \cite{CPBM06}
and
\emph{multi-environment P systems} 
\cite{Pardini-thesis}. In
both extensions, probabilities are associated with rules and the values may
vary
through time. A multi-environment P system model is
composed of a set of environment, each containing a probabilistic P system. All
the P systems inside the environments share some common features such as the
alphabet of symbols, membrane structure and evolution rules. Multi-environment P
systems have been used to model a simple (but realistic enough) ecosystem where a carnivore and several herbivorous species interact \cite{Colomer2012}.

Another class of term rewriting formalism, similar to P systems, is the
\emph{calculus of looping
sequences} (\textsc{cls}) that allows to represent membranes explicitly with
the calculus syntax. By the means of rewriting rules, it is possible to create,
dissolute and merge membranes \cite{Pardini-thesis}.
Recently, Pardini et al. extended this calculus with spatial information
\cite{BMMP09}.

\section{Other approaches}
There exists a variety of other  proposals which introduce locations or
compartments
into formal frameworks. In what follows we explain some of them. 

\paragraph{Processes in Space. } This modelling language \cite{Cardelli-CiE2010} has been used
to analyze the influence of network topologies on local and global dynamics of metapopulation systems ~\cite{three}.

\paragraph{Cellular Automata. }


\emph{Cellular automata} were proposed by Von Neumann, 
in 1949 . There are one-, two- and
three-dimensional cellular automata.
Cellular automata schematize the space into a regular
lattice, according to the
 spatial scale of the system studied. Each cell has some properties
and takes a value from a finite set of states. The value is updated in discrete time
steps. Rules are updated with a function that depends on the current state of
the cell
itself and its neighbored cells. 
The rules can be deterministic, stochastic or empirical. 
There are several tools for the simulation of cellular automata. 
Chen et al. developed a rule-based stochastic cellular automata model 
to simulate the
classical predator-prey Lotka-Volterra model
\cite{CYL09}.

\paragraph{Petri nets. }

Heiner et al. described a \emph{Petri net}-based framework for modelling and
analyzing
biochemical pathways \cite{HGD08}. The framework 
unifies qualitative, stochastic and continuous paradigms. According to
Heiner et al., each perspective adds
its contribution to the understanding of the system. Heiner et al. modeled a
\emph{extracellular signal-regulated kinases (ERK) 
transduction
pathway system} \cite{HGD08}. 
A pathway describes a sequence of interactions among
molecules \cite{Pardini-thesis}. In the particular case of
signal transduction pathway, there is a process, called \emph{receptor}, which
receives an external stimulus. The receptor triggers a chain of reaction which
end on some internal result.

Heiner et al. focused on transient behavior analysis of the transduction system.
 The stochastic Petri net describes a system
of stochastic reaction rate equations and the continuous Petri net is an
structured description of ordinary differential equations.
Note that arcs in Petri can be annotated with stoichiometric information.
 Qualitative Petri nets can be analyzed by a branching-time temporal logic. The
stochastic Petri nets can be analyzed by a continuous stochastic logic.  
As usual, the continuous model replaces the discrete values of species with
continuous
values which represent the overall behavior via concentrations.  This means
that the
concentrations of a particular species in such a model will have the same value
at each point of time for repeated experiments. 
Continuous
model can also be analyzed with a
linear-time temporal logic.

\paragraph{Grid Systems. }

\emph{Grid Systems} is a formalism to model population dynamics proposed by Barbuti et al. \cite{grid}. The formalism is inspired by concepts of
P systems and spatiality dynamics of cellular automata. In Grid Systems, environmental events that change population behaviour can
be defined as rewrite rules. There is a simulation tool for Grid Systems developed in Java \cite{grid}.

Grid Systems was used to model a population model of a species of mosquitoes,
\emph{Aedes albopictus}, that considers three types of external events: temperature
change, rainfall, and desiccation \cite{grid}. The events change the behaviour of the
species directly or indirectly. Each individual in the population can move
around in the ecosystem. 

\section{Conclusions}

In this article, we presented different formal languages for individual-based modelling of ecosystems.
Some are based on process calculi, some on P systems and others on cellular automata
or Petri nets. All these existing formal languages are complementary to each other and
should not be seen as competition.

We presented several stochastic formalisms, for instance, stochastic process calculi
and stochastic P systems. An advantage of stochastic formalisms is that
they have semantics in terms of continuous-time Markov chains. It is well-known that continuous-time
Markov chains for large numbers of individuals can be approximated by ordinary differential equations (\textsc{ode}) and 
there are several numeric methods that are computationally efficient to simulate \textsc{ode}s. The disadvantage
of these formalisms is that, sometimes, it is more intuitive to model systems with discrete time because  
stochastic rates are not known and, instead, we only know the probabilities that an individual will change from
one state to another.

We also presented several probabilistic formalisms, for instance, probabilistic process calculi
and probabilistic P systems. As we argued before, an advantage of this type of formalisms is that, sometimes,
it is easier to model systems in discrete time by considering the probabilities that an individual
will execute a certain action, instead of modelling stochastic rates. However, the semantics
of these languages suffer from state-space explosion, since they have an exponential number of
states. A solution for this problem is to represent the average behavior of the system using
mean-field equations, as it was done for \textsc{wsccs} and \textsc{s-palps}. Another disadvantage 
of these formalisms is that they usually consider both probabilistic and non-deterministic behavior. This
semantics limits the properties that can be verified for these systems because non-deterministic
behavior cannot be approximated by probabilities.

There is another interesting category of formalisms which are spatial extensions of probabilistic
and stochastic process calculi or P systems.  The semantics of these modelling languages is either
stochastic or probabilistic, but they offer simpler ways to represent space --either continuously or
discrete--, which facilitates the modelling of complex temporal properties of population systems.

Spatial, probabilistic and stochastic modelling languages have some drawbacks as modelling formalisms for ecosystems. Garavel argues that models based on
process calculi are not widespread because there are many calculi and many variants for each calculus, being difficult to choose the most appropriate \cite{garavel}. Garavel also argued that existing tools for process calculi are not user-friendly. We claim that these problems are also present in P systems and other formalisms to model population systems.

A research direction is how to develop user-friendly tools to simulate process calculi and P systems; for instance, to develop 
web applications or mobile applications that do no require extensive knowledge on Unix, Latex or programming to be used.
This could help to make these formalisms spread world wide.

Another research direction is to analyze and simulate larger case studies, studying how different evolutions can arise, depending on different
assumptions, such as spatial configurations, interaction rates and the role of the environment, high-level
reasoning about them.  Since spatial dynamics and temporal evolution of ecological systems are of key
interest, an interesting further step will be to develop a complementary spatio-temporal
logic, suitable for formally describing and verifying spatio-temporal properties of the different formalisms. 
A research on that direction is the three-valued spatio-temporal logic developed by Luisa Vissat et al. \cite{Vissat2017}.

Finally, we argue that research in formal modelling languages for population systems in ecology has produced a vast corpus of deep, valuable results; however, these results are only known, understood and applied by a small fraction of computer scientists. As a consequence, their practical impact is not as strong as it could be, in spite of successful attempts at using process calculi, P systems and other approaches to model and verify properties of ecological systems. We believe, however, that these formal modelling languages still have an important role to play in the future, but more interdisciplinary work is needed in collaboration with ecologists, software developers and policy makers.

\bibliographystyle{abbrv}

\end{document}

%% file: macros.tex
\long\def\symbolfootnote[#1]#2{\begingroup%
\def\thefootnote{\fnsymbol{footnote}}\footnote[#1]{#2}\endgroup}

\newcommand{\ccs}{\textsc{ccs}}
\newcommand{\ccp}{\textsc{ccp}}
\newcommand{\csp}{\textsc{csp}}

\newcommand{\prism}{\textsc{prism}}
\newcommand{\palps}{\textsc{palps}}
\newcommand{\spalps}{\textsc{s-palps}}
\newcommand{\wsccs}{\textsc{wsccs}}
\newcommand{\sccs}{\textsc{sccs}}
\newcommand{\pepa}{\textsc{pepa}}

\newcommand{\comment}[1]{}

